# QUANTUM FLUCTUATION THEOREMS AND WORK–ENERGY RELATIONSHIPS WITH DUE REGARD FOR CONVERGENCE, DISSIPATION AND IRREVERSIBILITY

Carolyne M. Van Vliet[*]

*Department of Physics, University of Miami,
P.O. Box 248046, Coral Gables FL 33124-0530, USA*

**Abstract.** Firstly the fluctuation theorems (FT) for *expended work* in a driven nonequilibrium system, isolated or thermostatted, together with the ensuing Jarzynski work–energy (W–E) relationships, will be discussed and reobtained. Secondly, the fluctuation theorems for *entropy flow* will be reconsidered. Our treatment will be fully quantum-statistical, being an extension of our previous research reported in Phys. Rev. E (2012), and will avoid the deficiencies that afflicted previous works such as: arguments based on classical trajectories in phase space, a reliance on the 'pure' von Neumann equation or 'non-reduced' Heisenberg operators, or other departures from the general tenets spelled out by Lindblad and others (e.g. Breuer and Petruccione) such as stochastic 'jump-induced' random trajectories. While a number of relationships from such previous works will still be employed, our Markov probability $P(\sigma_f, t_f | \sigma_0, t_0)$ shall only denote the two state-points, with no reference whatsoever to stochastic trajectories, these being meaningless in a quantum description.

___________________________________________________________________________________

## Contents

___________________________________________________________________________________

## 1. Introduction

### 1.1  General considerations

Thermodynamic systems, driven far from equilibrium by some protocol $\xi(t)$, operating over a time interval $t_0 \to t'$, have dominated the literature on nonequilibrium statistical mechanics since the early nineteen-nineties. While thermodynamic inequalities for such processes have been known since the days of Clausius, the new endeavour aimed at obtaining precise results, denoted as *Fluctuation Theorems* (FT), which compare the probabilities for certain 'action integrals', like work, entropy production, etc., for normal (or forward) processes *versus* the probability in reverse (or backward) processes — not forbidden by the second law of thermodynamics, but subject to the requirement that the final state for the process satisfy the historic inequalities referred to above, providing measurements are made after re-equilibration of the system. Generally, both closed (or isolated) and open (usually thermostatted or isothermal) systems have been considered.

---

[*] This article was submitted posthumously with edits by the author's grandson Dr. David S. Sukhdeo (dsukhdeo@alumni.stanford.edu)

Our ideas in this article have been largely shaped by our extensive earlier work on Linear Response Theory (LRT) [1-3], [4]. To obtain the Kubo–Green relations, it is essential that a general canonical ensemble be employed, such as the canonical, grand-canonical or pressure ensemble; the associated state functions are the Helmholtz free energy $F$, the grand potential $\Omega$ and the Gibbs free energy $G$, respectively. A special class of FT's pertains to isothermal driven systems that can *instantly dispense heat* to the thermal bath; the transfer process is succinctly described in two early papers by Crooks [5, 6]. For such a bath, the entropy change is the 'dissipative work', $T\Delta S = w - \Delta F^0 \equiv w_{diss}$, (as is found from the first and second Law when $\Delta Q = 0$), where the superscript zero refers to the equilibrium state function. Let now $p(w)$ denote the probability that work $w$ is performed in the forward process and $\tilde{p}(-w)$ the probability that work $-w$ is delivered in the reverse or backward process; the Crooks–Tasaki FT [5, 7] then reads

$$p(w) / \tilde{p}(-w) = \exp(\beta w_{diss}), \tag{1.1}$$

where $\beta^{-1} = k_B T^r$, $k_B$ being Boltzmann's constant and $T^r$ the reservoir temperature. If other reservoirs are present, we need to redefine the dissipative work. If the volume is variable as in the pressure ensemble we have $T\Delta S = w - \Delta G^0 \equiv w_{diss}$, while for a grand-canonical ensemble $T\Delta S = w - \Delta \Omega^0 \equiv w_{diss}$. With these extensions, Eq. (1.1) contains three forms of *expended work* FT's.

It is well-known that FT's like (1.1) have an associated *work–(free) energy* (W–E) relationship, obtained by slight rewriting of (1.1); for the case that the only reservoir is the thermal bath, the form $p(w)\exp(-\beta w) = \tilde{p}(-w)\exp(-\beta \Delta F^0)$ is integrated over all $w$; we note that from a stochastic viewpoint this is a *two-point* integration. In either case, the result is the W–E relationship

$$\langle e^{-\beta w} \rangle = e^{-\beta \Delta F^0}, \tag{1.2}$$

first obtained by Christopher Jarzynski [8], actually prior to the Crooks–Tasaki FT. Since by Jensen's inequality, $e^{\langle x \rangle} \leq \langle e^x \rangle$, *a posteriori* this confirms the Clausius result for isothermal open systems, $\langle w \rangle \geq \Delta F^0$. We shall not reference here the dozens of derivations that have been given in the literature in the decade after the above results were first published (1997/98 – 2008), both by the authors themselves and by scores of others, except for a really elegant derivation by Talkner and Hänggi in 2007 [9], on which we comment later (Section 2.1). Many references can be found in our own article on *time-reversal symmetry* in Phys. Rev. E of 2012 [10].

Usually little is said about the nature of the driving protocol $\xi(t)$. Since our purpose is to provide a quantum description for FT's and ensuing W–E relationships, we should not be guided by thermodynamics, but consider the microscopic formulation of LRT. Kubo theory commences by adding an external 'response Hamiltonian', $AF(t)$, in the von Neumann equation; here A is a system operator and $F(t)$ is a generalized external field, composed of *c*-numbers. A microscopic treatment is only possible if the protocol is associated with the external field, i.e. $\xi(t) \Rightarrow F(t)$. Since the processes envisaged are nonstationary, the response Hamiltonian is now $A(t)F(t)$; the solution of the new Kubo-type von Neumann equation is easily obtained.[1] If, on the other hand, the protocol involves variation of the system operator A (or a set of operators $A_i$), which generally do not commute with the Hamiltonian H of the system, we must resort to coarse-graining of the eigenstates of the pertinent variables in '*a-space*', as discussed by van Kampen [11] and resulting in a *mesoscopic* treatment. This was considered in detail in a previous paper; see Van Vliet [12].

As a first example, already mentioned by Crooks [6], let us consider an Ising spin lattice that is being magnetized (forward process) or demagnetized (backward process) by a field $H(t)$, the response Hamiltonian being $\sum_i \hat{\mu}_i(t) H(t) \propto \sum_i \hat{\sigma}_i(t) H(t)$, where the $\{\hat{\sigma}_i\}$ are the spin operators. The process is discrete and composed of successive individual spin-flips. The environment is held at a constant temperature $T^r$. For the work FT and the Jarzynski relationship to be applicable for this microscopic system, the spin relaxation time must be near zero in order that heat is instantly exchanged. This is a tall order that can only be met in computer simulations. If any finite relaxation occurs, spin-flips will overlap and a range of spins must be bundled to form mesoscopic state-variables $\{\sigma_\kappa\}$ as we will see more clearly when dealing with the entropy FT's in Section 3.2; (the entropy flow, as defined there, is zero only for the single spin-flip microscopic case).

In another example, found in both of Crooks papers [5,6] as well as in a later article by Cohen and Imry [13], a gas is compressed or decompressed by a piston in a cylinder with a diathermal wall and embedded in a heat bath. The driving force is obviously related to the pressure which, in turn, depends on the classical virial [14]. To simplify the discussion, let the gas that is compressed be an ideal gas. Because of its confinement, the gas is essentially one-dimensional and its external virial is $\sum_i x_i (\partial H/\partial x_i)$, where the $\{x_i\}$ denote the positions of the molecules in the direction of the displacement. For the derivative it suffices to mark the position of one selected molecule, labelling its position $\xi$, where $\xi(t)$ monitors the position of the piston. This suggests that the response Hamiltonian should be given by $\sum_i x_i(t) \partial H[\xi(t)]/\partial \xi$. Elementary models can be constructed so that the derivative $\partial H/\partial \xi$ is not entirely singular (cf. [13] Section V). However, we now deal with a variation of an operator that does not commute with the Hamiltonian. Thus, contrary to the presumptions in [13], there is no microscopic quantum description for this classical textbook case; the eigenstates $\{a_I\}$ pertain to the coarse-grained operators involved and the behaviour is governed by the mesoscopic Master equation in *a*-space, with the FT's and W–E relationships as found in Ref. [12] and further discussed in Section 2.2.

We now briefly review the various entropy production FT's. These were actually formulated prior to the expended work FT's discussed above. The first article on the subject stems from Evans, Cohen and Morriss [15], dealing with violations of the Second Law for the shear stress in fluids driven far from equilibrium. The results were re-examined in a more general way by many researchers, among others Evans and Searles [16], Gallavotti and Cohen [17, 18], Kurchan [19, 20], Lebowitz and Spohn, [21], Maes [22], Harris and Schütz [23], and Seifert [24]. Anyone studying these developments — the articles, [20] excepted, are listed in historical order — notices that, whereas all studies deal with classical phase space trajectories, the first papers, refs. [15-18], assume certain *chaoticity rules* for the trajectories to be obeyed, while refs. [19-23] deal with ordinary *stochastic paths* employing a Langevin, Kramers, or Master equation approach. Since in this paper we envision to obtain the quantum entropy FT's, we expected to find a classical principle that would emerge in the correspondence limit. Indeed, Maes wrote a (quite abstract) article on entropy FT's as a Gibbs' property, introducing his "Gibbs Measure". While he defends the possibility of using the chaotic hypothesis: "a reversible many-particle system in a stationary state can be regarded as a transitive Anosov system for the purpose of computing the macroscopic properties", he leaves us the benefit of the doubt as to whether the FT's based on chaotic trajectories should necessarily apply to, or be identical with, FT's for entropy production associated with regular stochastic trajectories; perhaps we glean an insight into his reasoning when in his concluding remarks he recommends that studies be done for information-entropy FT's.

---

[1] While we could pursue a new Kubo solution with generalized Heisenberg operators, there is no real need for it since these Heisenberg operators behave similarly in the dual space as the density operator in the direct Hilbert space. So, the results for the reduced correlation functions will be the same as obtained from the new nonstationary Master equation.

It may be useful to outline the original derivation employing the rules for chaotic systems, cf. [16]. In an equilibrium state the volume in phase space is a Poincaré invariant. Ruelle [25] has shown that in a nonequilibrium steady state the volume contracts. Let us consider a trajectory that is cut in segments of duration $\tau$ labelled '$i$'. The track is a transformation $\xi \to \varphi(\xi)$, $\xi \in \Delta\Omega$, where we assume that in some sense $\Delta\Omega$ is a bounded manifold of dimension $2rN$. After a time $\tau$ the new manifold $\Delta\Omega'$ is a diffeomorphism of $\Delta\Omega$. Let $\lambda_n^\pm$ be the spectrum of Lyapunov exponents. The normalized invariant measure for a multidimensional system is given by

$$\mu_i(\tau) = \frac{\Lambda_i^{-1}}{\sum_i \Lambda_i^{-1}} = \frac{\exp[-\sum_n \lambda_{n,i}^+ \tau]}{\sum_i \exp[-\sum_n \lambda_{n,i}^+ \tau]}, \quad (1.3)$$

where $\Lambda_i$ is the product of all expanding eigenvalues of the stability matrix associated with the Hessian and $\{\lambda_{n,i}^+\}$ the set of positive local Lyapunov exponents on segment '$i$'. Let now $\tilde{\mu}_i(\tau)$ be the measure for a time-reversed path. Since the Hamiltonian is even in the momenta (providing the magnetic field is reversed upon time reversal), we have the time-reversal property $\tilde{\lambda}_{n,i}^\pm = \lambda_{n,i}^\mp$. Hence, we have

$$\frac{\tilde{\mu}_i(-\tau)}{\mu_i(\tau)} = \frac{\exp[\sum_n \tilde{\lambda}_{n,i}^+ \tau]}{\exp[-\sum_n \lambda_{n,i}^+ \tau]} = \frac{\exp[\sum_n \lambda_{n,i}^- \tau]}{\exp[-\sum_n \lambda_{n,i}^+ \tau]}$$

$$= \exp[\sum_n (\lambda_{n,i}^+ + \lambda_{n,i}^-)\tau] = \exp(-rN\langle\alpha\rangle_i \tau). \quad (1.4)$$

The last equality is based on the *pairing property* for Lyapunov exponents discussed by Gallavotti and Cohen. Note that the sum is zero in a volume-preserving equilibrium system; thus, for a far from equilibrium volume-contracting system, $\langle\alpha\rangle$ is a measure for the entropy production rate, which later (Section 3.2) will be denoted by $\hat{\eta}$. If $\Delta S$ (in units $k_B$) denotes the microscopic entropy change associated with the paths segments of duration $|\tau|$, then (1.4) leads to the FT:

$$\tilde{p}_\tau(-\Delta S)/p_\tau(\Delta S) = e^{-\hat{\eta}\tau}. \tag{1.5}$$

Let now at some point in time the system reach a steady state, from whereon $\tilde{p} = p$ and $\Delta S = \hat{\eta}\tau$. This leads to the asymptotic statement $p_\tau(-\hat{\eta}) = p_\tau(\hat{\eta})e^{-\eta\tau} \sim 0$: second law violating intervals vanish exponentially with time and for macroscopic time die out, cf. Evans and Searles [16]. Or, stated differently,

$$\lim_{\tau\to\infty}\left(-\frac{1}{\tau}\ln\left[\frac{p_\tau(-\hat{\eta})}{p_\tau(\hat{\eta})}\right]\right) = \hat{\eta}, \tag{1.6}$$

which is the form implied in most treatments, cf. Kurchan [19] and Lebowitz and Spohn [21].

### 1.2 Deficiencies of previous work

(i). All of the above cited papers — with the exception of Tasaki [7], Talkner and Hänggi [9], Cohen and Imry [13] and Kurchan [20] — are explicitly based on a classical phase space description, which apparently enjoyed a (possibly undeserved) rejuvenation around the turn of the century, both in the physical and mathematical literature. While quantum mechanics can be formulated in phase space using the Wigner distribution [26, 27], this distribution is not positive definite, unless integrated over a volume $h^{rN}$; this permits at most the concept of "*fuzzy*" trajectories.

(ii) The quantum treatments in the 'excepted' articles listed above are based on the 'pure' von Neumann equation or the 'non-reduced' Heisenberg operators whose solutions do not converge. Let us assume that local derivatives [ $\partial\rho/\partial t$, $\partial A_H/\partial t$ ] are zero; here $\rho$ is the density operator and $A_H$ is a Heisenberg operator. The Hamiltonian is time-dependent, giving rise to the evolution operator

$$U(t_f,t_0) = \mathrm{T}\,\exp\left[-(i/\hbar)\int_{t_0}^{t_f}\mathrm{H}(\vartheta)d\vartheta\right], \tag{1.7}$$

where T is the time-ordering operator. Generally, the time dependence of $\mathrm{H}[\xi(t)]$ is parametrized by introducing a set of time points $t_i$ $(i = 0,1...,n)$ with $t_n = t_f$; on an interval $(t_k \leq t < t_{k+1})$ the Hamiltonian is assumed to be constant and is denoted by $\mathrm{H}_{\xi_k}$; cf. Fig. 1. The evolution operator then reads:

$$U(t_f, t_0) = \text{T} \exp[-(i/\hbar)\sum_{i=0}^{n-1} \text{H}_{\xi_i}(t_{i+1} - t_i)]$$

$$= \prod_{i=0}^{n-1} \exp[-(i/\hbar) \text{H}_{\xi_i}(t_{i+1} - t_i)] = \prod_{i=0}^{n-1} U_{\xi_i}(t_{i+1}, t_i). \quad (1.8)$$

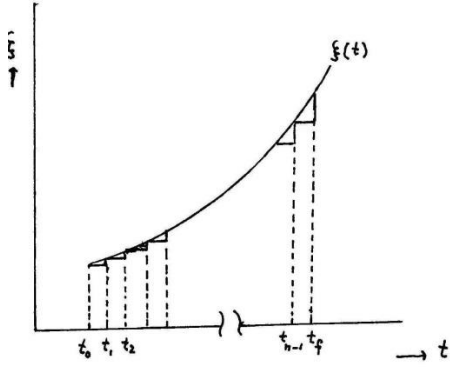

Fig. 1. Segmentation of the protocol $\xi(t)$.

The solutions of the 'pure' von Neumann equation and of the 'non-reduced' Heisenberg equation of motion can now be written in terms of products of stationary evolution operators; or more succinctly by employing the Liouville superoperator in the Liouville space $\mathbf{S} \otimes \tilde{\mathbf{S}}$:

$$\exp(-i L_\xi t) = \{\exp[-(i/\hbar) \text{H}_\xi t] \rightarrow \quad \leftarrow \exp[(i/\hbar) \text{H}_\xi t]\}. \quad (1.9)$$

We thus have

$$\left. \begin{array}{l} \rho(t) \\ A_H(t) \end{array} \right\} = \{\prod_{i=0}^{n-1} \exp(\mp i(t_{i+1} - t_i) L_{\xi_i})\} \left\{ \begin{array}{l} \rho(t_0) \\ A_H(t_0) \end{array} \right.. \quad (1.10)$$

This is nothing than a rotation in the Liouville space with divergent resolvent, cf. Fano [28].

(iii) Dissipation must be accounted for by considering interactions, $\text{H} = \text{H}^0 + \lambda V$, to all orders of perturbation. This yields Lindblad's quantum master equation (QME) [29, 30] and a similar result for the reduced Heisenberg equation; cf. also Breuer and Petruccione [31]). Confining ourselves to stationary evolution, the result for the density operator is $\rho(t) = \exp(tL^{r\dagger})\rho(t_0)$, with

$$L^{r\dagger}(\rho) = \sum_i \{-\tfrac{1}{2}[V_i^\dagger V_i, \rho]_+ + V_i \rho V_i^\dagger\} - (i/\hbar)[\text{H}^1, \rho], \quad (1.11)$$

where $[..,..]_+$ denotes the anticommutator and where the $\{V_i\}$ are operators in $\mathbf{S}$. The Hamiltonian $\text{H}^1$ is a dressed form of $\text{H}^0$ and reduces to it when the interactions are weak. In Ref. [10] we computed the weak coupling, long time limit for the von Neumann equation, arriving at the Pauli–Van Hove ME (cf. [32]), the result being concordant with the general Lindblad QME. The reduced results for (1.10), leaving out the segmentation of the interval, are summarised by:

$$\lim_{\lambda, t} e^{\pm itL} = e^{-t(\Lambda_d \mp i L^0)}, \quad (1.12)$$

where $L^0 K = (1/\hbar)[\text{H}^0, K]$ is the zero-order Liouville operator associated with $\text{H}^0$; note that this is just the last term in the Lindblad QME (1.11). The dissipation stems, however, fully from the real *positive semi-definite* master operator $\Lambda_d$ in diagonal Liouville space [1, Section 8.1], to wit:

$$\Lambda_d K = \sum_\gamma |\gamma\rangle\langle\gamma| M_\gamma[\langle\gamma|K|\gamma\rangle], \tag{1.13}$$

where $M_\gamma$ is the ordinary master operator in function space,

$$M_\gamma[f(\gamma)] = -\sum_{\bar{\gamma}} \{W(\gamma|\bar{\gamma})f(\bar{\gamma}) - W(\bar{\gamma}|\gamma)f(\gamma)\}. \tag{1.14}$$

Here the $\{|\gamma\rangle\}$ are the eigenstates of $H^0$ and the $W$'s are the transition rates; by Fermi's golden rule

$$W(\bar{\gamma}|\gamma) = (2\pi\lambda^2/\hbar)|\langle\gamma|V|\bar{\gamma}\rangle|^2 \delta(E_\gamma - E_{\bar{\gamma}}) = W(\gamma|\bar{\gamma}). \tag{1.15}$$

In this article the segmentation method will not explicitly be called upon except in subsection 3.2; rather, for the nonstationary case a time-dependent master superoperator $\Lambda_d(t)$ will be employed.

## 2. Quantum Fluctuation Theorems for expended work and Work–Energy relationships

### 2.1 Results from the microscopic ME for systems that admit a microscopic treatment

The previous section showed clearly that all FT's depend on a joint consideration of forward and reverse processes. In a theory without perturbations time reversal symmetry follows directly from the evolution operator. Denoting the eigenstates (ES) of $H_i^0$ by $\{|\gamma_i\rangle\}$, the diagonal part of the density operator $p(\gamma_i, t) = \langle\gamma_i|\rho(t)|\gamma_i\rangle$ for $t > t'$ is found from

$$\begin{aligned} p(\gamma, t) &= \langle\gamma|U(t, t')\rho(t')U^\dagger(t, t')|\gamma\rangle \\ &= \sum_{\gamma'\gamma''} \langle\gamma|U(t, t')|\gamma'\rangle\langle\gamma'|\rho(t')|\gamma''\rangle\langle\gamma''|U^\dagger(t, t')|\gamma\rangle, \end{aligned} \tag{2.1}$$

where $U$ is given by (1.7) and we inserted the decomposition of unity $\Sigma_\gamma |\gamma\rangle\langle\gamma| = \mathbf{1}$. With an initial random phase assumption $\langle\gamma'|\rho(t')|\gamma''\rangle = p(\gamma', t')\delta_{\gamma'\gamma''}$, we obtain

$$p(\gamma, t) = \sum_{\gamma'} |\langle\gamma|U(t, t')|\gamma'\rangle|^2 p(\gamma', t'). \tag{2.2}$$

From Bayes' rule we note that this implies a nonstationary conditional probability $P$,

$$P_{\text{ns}}(\gamma, t|\gamma', t') = |\langle\gamma|U(t, t')|\gamma'\rangle|^2. \tag{2.3}$$

Now,

$$\begin{aligned} \langle\gamma|U(t, t')|\gamma'\rangle^* &= \langle\gamma'|U^\dagger(t, t')|\gamma\rangle = \langle\gamma'|\tilde{U}(t', t)|\gamma\rangle, \\ \langle\gamma|U(t, t')|\gamma'\rangle &= [\langle\gamma'|U^\dagger(t, t')|\gamma\rangle]^* = [\langle\gamma'|\tilde{U}(t', t)|\gamma\rangle]^*. \end{aligned} \tag{2.4}$$

where $\tilde{U}(t', t)$ is the evolution operator for the backward process with inverse time ordering. Thus multiplying the two statements, we obtain

$$P_{\text{ns}}(\gamma, t|\gamma', t') = \tilde{P}_{\text{ns}}(\gamma', t'|\gamma, t). \tag{2.5}$$

Equation (2.5) expresses the time-reversal symmetry. [Not that *all* processes possess this symmetry; cf. Wang and Feldman [33].]

This has, however, been a rather futile exercise; we need perturbations of 'the motion proper', just as Boltzmann considered collisions that perturbed the streaming motion due to ponderomotive gradients and fields. In order to find the reduced density operator or the reduced Heisenberg operators, the full perturbation procedure was carried out in our article [10] and many other places. For the reduced evolution operator, cf. Eq. (1.12), we established there with much effort

$$\langle \gamma | U^R(t,t') | \gamma' \rangle^* = \langle \gamma' | \tilde{U}^R(t',t) | \gamma \rangle. \qquad (2.6)$$

Multiplying both sides with their complex conjugates, we once more find

$$P^R_{\text{ns}}(\gamma,t | \gamma',t') = \tilde{P}^R_{\text{ns}}(\gamma',t' | \gamma,t). \qquad (2.7)$$

Now let us change the interval notation with $(t',t) \Rightarrow (t_0, t_f)$. We multiply the left-hand side with $p_{\text{can}}(\gamma_0,t_0)$ and the right-hand side with $p_{\text{can}}(\gamma_f,t_f)$ to obtain,

$$P_{\text{ns}}(\gamma_f,t_f | \gamma_0,t_0) p_{\text{can}}(\gamma_0,t_0) = \tilde{P}_{\text{ns}}(\gamma_0,t_0 | \gamma_f,t_f) p_{\text{can}}(\gamma_f,t_f)[p_{\text{can}}(\gamma_0,t_0) / p_{\text{can}}(\gamma_f,t_f)], \qquad (2.8)$$

whereby it is understood that the ES $|\gamma_f\rangle$ and $|\gamma_0\rangle$ belong to different Hamiltonians. With $p_{\text{can}}$ given by the Gibbs distribution, $(1/Z)e^{-\beta \varepsilon_\gamma}$, we have by Bayes' rule [34]

$$W_2(\gamma_f,t_f;\gamma_0,t_0)e^{-\beta(\varepsilon_{\gamma_f}-\varepsilon_{\gamma_0})} = \tilde{W}_2(\gamma_0,t_0;\gamma_f,t_f)e^{-\beta(F_f^0-F_0^0)}; \qquad (2.9)$$

here $W_2$ is the two-point probability distribution and $F^0$ denotes the equilibrium Helmholtz free energy of the system. Or also, with $\Delta F^0 = F_f^0 - F_0^0$,

$$W_2(\gamma_f,t_f;\gamma_0,t_0)e^{-\beta(\varepsilon_{\gamma_f}-\varepsilon_{\gamma_0})} = \tilde{W}_2(\gamma_0,t_0;\gamma_f,t_f)e^{-\beta \Delta F^0}; \qquad (2.10)$$

From this the Crooks–Tasaki FT will be obtained below.

*The operator for work*

We now come to the notion of work in quantum mechanics; many essays having been written on this concept. Allahverdyan and Nieuwenhuizen associated the work operator with the Heisenberg Hamiltonian difference [35]

$$\mathrm{H}_H(t_f) - \mathrm{H}_H(t_0) = \int_{t_0}^{t_f} \left(\partial \mathrm{H}_H(s)/\partial s\right) ds = \int_{t_0}^{t_f} \left(\partial \mathrm{H}_H(s)/\partial s\right) \dot{s}\, dt, \qquad (2.11)$$

where we note that $d\mathrm{H}_H/dt = \partial \mathrm{H}_H/\partial t$ because of the Heisenberg equation of motion and $\mathrm{H}_H(t_0) = \mathrm{H}(t_0)$. This operator is, however, subject to strong fluctuations [36]. Consequently, various authors, among whom Talkner et al. [37] concluded that "work is not an observable". We beg to differ, noting that an operator for work, denoted by $\Omega$ does exist, but is not in the Hilbert space $\mathbf{S}^0$ of $H^0$. However, as noted by Johann von Neumann [38], one can construct a unitary operator in $\mathbf{S}^0$ that by Stone's theorem has a spectral resolution [39, 40]

$$e^{iuW} = \int e^{iuw} d\hat{p}(w), \qquad (2.12)$$

where $\{w\}$ are the EV and $d\hat{p}(w)$ are the incremental projectors. Let $\rho$ be the appropriate density operator, then (2.12) yields the characteristic function[2]

$$X(u) \equiv \langle e^{iuW} \rangle = \text{Tr}\left(\rho \int e^{iuw} |w\rangle\langle w| dw\right)$$

$$= \int e^{iuw} \langle w|\rho|w\rangle dw = \int e^{iuw} p(w) dw; \qquad (2.13)$$

here $p(w)$ is the probability density function, which may be singular[3]. Contrary to our incorrect story in [10], we shall now use the reduced Heisenberg operators, writing [NB. 'X' is cap. chi],

$$X(u) = \langle T\{e^{iu[H_H^R(t_f) - H_H^R(t_0)]}\}\rangle$$

$$= \langle T\{e^{iuH_H^R(t_f)} e^{-iuH(t_0)}\}\rangle, \qquad (2.14)$$

where non-subscripted operators are Schrödinger operators.

*From the Heisenberg picture to the Schrödinger picture*

Let us now consider the nonstationary Heisenberg correlation function for any two operators $A_H^R(t_f)$ and $B(t_0)$, which may belong to different Hilbert spaces, $\mathbf{S}[H_f^0]$ and $\mathbf{S}[H_0^0]$. It will be expedient to consider averages in the tensor product space $\mathbf{S}[H_f^0] \otimes \mathbf{S}[H_0^0]$ with ES $\{|\gamma_f \gamma_0\rangle\}$; so,

$$\Phi_{AB}(t_f, t_0) = \tfrac{1}{2}\text{Tr}\{\rho_{f,0}[A_H^R(t_f), B(t_0)]_+\}$$

$$= \tfrac{1}{2}\text{Tr}\{\rho_{f,0}[A_{Hd}^R(t_f), B_d(t_0)]_+\} = \text{Tr}\{\rho_{f,0}[A_{Hd}^R(t_f) B_d(t_0)]\}. \qquad (2.15)$$

Here, $[..,..]_+$ denotes the anticommutator; the operators have been split into a diagonal and non-diagonal part, whereby only the diagonal part contributes to dissipation as noted before — whence the final right-hand side. Next, we need the ME in Liouville space, derived in [10, Eq. (3.15)] for the stationary case. Since in [10] the nonstationary case was handled via the segmentation procedure, the final solution for the nonstationary density operator is readily written down, cf. the result [10, Eq. (3.22)], [3]

$$\rho_{ns}(t_f) = \left\{T \exp\left[-\int_{t_0}^{t_f} d\vartheta \Lambda_d(\vartheta)\right]\right\} \rho_{ns}(\vartheta). \qquad (2.16)$$

We now turn to the Heisenberg operator correlation form (2.15). We saw previously that the diagonal parts of the reduced operators have the same time behaviour as the density operator, cf. Eq. (1.12). Hence, we have

$$\Phi_{AB}(t_f, t_0) = \text{Tr}\left\{\rho\, T \exp\left[-\int_{t_0}^{t_f} d\vartheta\, \Lambda_d(\vartheta)\right] A_d(\vartheta) B_d(t_0)\right\}. \tag{2.17}$$

---

[2] Justification needs the standard mathematical notation for the scalar product and the projectors; the rhs then yields the Lebesgue–Stieltjes integral $\int \exp(iuw) dP(w)$, where $P(w)$ is the cumulative distribution function (cdf).

[3] In Ref. [10] the final closed form was never written down but clearly implied, for in the end the segmentation in $n$ stretches should be subject to $n \to \infty$ leading to (2.16), a form akin to the original evolution operator in Eq. (1.7).

In the Heisenberg picture $\rho$ is constant, equal to $\rho_{\text{can}}$ if the system is thermostatted. *Formally* we work in the tensor product space but in practise this gives no complications, since for the matrix elements in the representation $\{|\gamma_f\, \gamma_0\rangle\}$ we have (see also, [10 Section IIA2]),

$$\begin{aligned}\langle \gamma_0 \gamma_f | A^R_{Hd}(\vartheta) | \gamma_f \gamma_0 \rangle &= \langle \gamma_f | A^R_{Hd}(\vartheta) | \gamma_f \rangle, \\ \langle \gamma_0 \gamma_f | B_d(t_0) | \gamma_f \gamma_0 \rangle &= \langle \gamma_0 | B_d(t_0) | \gamma_0 \rangle,\end{aligned} \tag{2.18}$$

since $A_d$ and $B_d$ belong to the different state spaces, $\mathsf{S}[H_f]$ and $\mathsf{S}[H_0]$, respectively. Hence we get, mindful of the decomposition (1.13) of the operator $\Lambda_d$, [4]

$$\Phi_{AB}(t_f, t_0) = \sum_{\gamma\gamma_0} p(\gamma_0) \left\{ T\exp\left[-\int_{t_0}^{t_f} d\vartheta\, M_\gamma(\vartheta)\right]\right\} \langle \gamma | A_d(\vartheta) | \gamma \rangle \langle \gamma_0 | B_d(t_0) | \gamma_0 \rangle. \tag{2.19}$$

This will be evaluated by a Green's function procedure. For a less cumbersome treatment, we shall at first work in an interval $(t' \to t)$. Thus, let us more closely look at

$$\langle \gamma | A^R_{Hd}(t) | \gamma \rangle = \left\{ T\exp\left[-\int_{t'}^{t} d\vartheta\, M_\gamma(\vartheta)\right]\right\} \langle \gamma | A_d(\vartheta) | \gamma \rangle. \tag{2.20}$$

A more expedient expression is obtained by differentiating the above, resulting in

$$\frac{\partial}{\partial t}\langle \gamma | A^R_{Hd}(t) | \gamma \rangle + M_\gamma(t)\langle \gamma | A^R_{Hd}(t) | \gamma \rangle = \delta(t^+)\langle \gamma | A_d(t') | \gamma \rangle. \tag{2.21}$$

[The right-hand side accounts for the initial condition at $t = 0$, as may be verified by Laplace transformation.] Let now $g(\gamma, t; \gamma', t')$ be the Green's function for the non-self-adjoint differential equation above; it satisfies

$$\frac{\partial g(\gamma, t; \gamma', t')}{\partial t} + M_\gamma(t) g(\gamma, t; \gamma', t') = \delta(t - t')\delta(\gamma - \gamma'). \tag{2.22}$$

The inhomogeneous equation (2.21) has the solution,[5] absent boundary terms from the bilinear concomitant, cf. Morse and Feshbach [41, Section 7.3]

$$\begin{aligned}\langle \gamma | A^R_{Hd}(t) | \gamma \rangle &= \int_0^{t+0} dt'\, \delta(t'^+) \int_{D(\gamma')} \Delta\gamma'\, g(\gamma, t; \gamma', t')\langle \gamma' | A_d(t') | \gamma' \rangle \\ &= \int \Delta\gamma'\, g(\gamma, t; \gamma', 0)\langle \gamma' | A_d | \gamma' \rangle.\end{aligned} \tag{2.23}$$

---

[4] Note that $\exp[\sum_\gamma |\gamma\rangle\langle\gamma| f(\gamma)] = \sum_\gamma |\gamma\rangle\langle\gamma| \exp[f(\gamma)]$, as is found by series expansion.

[5] Since the EV $\gamma$ are dense, one could work with probability density functions (pdf). However, we prefer to work with distributions, so we shall incorporate the density of states $Z(\gamma)$ in an interval $\Delta\gamma = Z(\gamma)d\gamma$; thus $\sum_\gamma \to \int \Delta\gamma$.

Presently, let us reconsider the ME for the nonstationary Markov probability $P_{\text{ns}}(\gamma,t|\gamma',t')$,

$$\frac{\partial P_{\text{ns}}(\gamma,t|\gamma',t')}{\partial t} + M_\gamma(t)P_{\text{ns}}(\gamma,t|\gamma't') = \delta(t-t')\delta(\gamma-\gamma'), \qquad (2.24)$$

where we included the initial conditions. Comparing with (2.22), we note the similarity, but since the delta function is even, $\gamma$ and $\gamma'$ could be reversed. Indeed, in [1, Eqs. (7.5), (7.27)] we showed

$$P_{\text{ns}}(\gamma,t|\gamma',t') = g(\gamma',t;\gamma,t'). \qquad (2.25)$$

Substituting (2.25) into (2.23) we obtained alternately,

$$\langle\gamma|A^R_{Hd}(t)|\gamma\rangle = \int \Delta\gamma' P_{\text{ns}}(\gamma',t|\gamma,0)\langle\gamma'|A_d|\gamma'\rangle. \qquad (2.26)$$

This, then, yields

$$\Phi_{AB}(t,0) = \iint \Delta\gamma\,\Delta\gamma' P_{\text{ns}}(\gamma',t|\gamma,0)p(\gamma)\langle\gamma'|A_d|\gamma'\rangle\langle\gamma|B_d|\gamma\rangle$$

or, upon interchange of $\gamma \leftrightarrow \gamma'$:

$$\Phi_{AB}(t,0) = \iint \Delta\gamma\,\Delta\gamma' P_{\text{ns}}(\gamma,t|\gamma',0)p(\gamma')\langle\gamma|A_d|\gamma\rangle\langle\gamma'|B_d|\gamma'\rangle. \qquad \mathbf{(2.27)}$$

Finally, going back to the original interval of interest, we established:

$$\Phi_{AB}(t_f,t_0) = \iint \Delta\gamma_0\,\Delta\gamma_f W_2(\gamma_f,t_f;\gamma_0,t_0)\langle\gamma_f|A_d|\gamma_f\rangle\langle\gamma_0|B_d|\gamma_0\rangle. \qquad \mathbf{(2.28)}$$

*This is the correlation function in the Schrödinger form* — the operators are constant while the time dependence is vested in $P_{\text{ns}}$ or $W_2$.

This will now be applied to the characteristic function for the work given in (2.14). One finds,

$$X(u) = \langle e^{iuH^R_H(t_f)}e^{-iuH^R_H(t_0)}\rangle = \iint \Delta\gamma_0\,\Delta\gamma_f\,W_2(\gamma_f,t_f;\gamma_0,t_0)\langle\gamma_f|e^{iuH(t_f)}|\gamma_f\rangle\langle\gamma_0|e^{-iuH(t_0)}|\gamma_0\rangle$$

$$= \iint \Delta\gamma_0\,\Delta\gamma_f W_2(\gamma_f,t_f;\gamma_0,t_0)\,e^{iu\varepsilon_{\gamma_f}}e^{-iu\varepsilon_{\gamma_0}}$$

$$= \iint \Delta\gamma_0\,\Delta\gamma_f W_2(\gamma_f,t_f;\gamma_0,t_0)\,e^{iu(\varepsilon_{\gamma_f}-\varepsilon_{\gamma_0})}. \qquad (2.29)$$

We note again that the average entails a two-point process. Inversion of the characteristic function gives the probability for the eigenvalues of the work operator, i.e. that work $w$ is performed,

$$p(w) = \iint \Delta\gamma_0\,\Delta\gamma_f\,W_2(\gamma_f,t_f;\gamma_0,t_0)\delta[w-(\varepsilon_{\gamma_f}-\varepsilon_{\gamma_0})]. \qquad (2.30)$$

Clearly, the work goes into the excitations of the system. Curiously, the above result is identical with that found in the paper by Talkner et al. [37, their Eq. (10)]. As with the time-reversal property, theories with no dissipation — as Kubo's original LRT — usually give the 'right' results except that entropy production is nil. In this paper random aspects are incorporated *ab initio*.
Going back to Eq. (2.10), we multiply by the delta function of (2.30), to obtain,

$$W_2(\gamma_f, t_f; \gamma_0, t_0)\delta[w-(\varepsilon_{\gamma_f}-\varepsilon_{\gamma_0})] = \tilde{W}_2(\gamma_0, t_0; \gamma_f, t_f)\delta[-w-(\varepsilon_{\gamma_0}-\varepsilon_{\gamma_f})]e^{\beta(w-\Delta F^0)}, \quad (2.31)$$

where we used that the delta function is even. Integrating both sides and noting (2.30) and its analogue for reversed time, we established

$$p(w)/\tilde{p}(-w) = e^{\beta(w-\Delta F^0)}, \quad (2.32)$$

which is the quantum Crooks–Tasaki fluctuation theorem. Next, we notice that for complex $u$, the domain of analyticity is $0 \leq \text{Im}\, u \leq \beta$, cf. [20]. Thus, setting $u = i\beta$ in (2.29) we find

$$\langle e^{-\beta W}\rangle = \iint \Delta\gamma\, \Delta\gamma'\, W_2(\gamma_f, t_f; \gamma_0, t_0)e^{-\beta w}, \quad (2.33)$$

as expected. Lastly by rearranging (2.32) as $p(w)e^{-\beta w} = \tilde{p}(-w)e^{-\beta\Delta F^0}$, integration over $w$ yields the quantum Jarzynski W–E theorem

$$\langle e^{-\beta W}\rangle = e^{-\beta\Delta F^0}. \quad (2.34)$$

*Talkner–Hänggi relation for the characteristic function*

By manipulation in the complex plane, Talkner and Hänggi [9] obtained a relation linking the characteristic functions $\text{X}(z)$ and $\tilde{\text{X}}(z)$ of the forward and backward processes, respectively, with $\text{X}(z) = \langle \exp(iz\text{W})\rangle_>$ and $\tilde{\text{X}}(z) = \langle \exp(iz\text{W})\rangle_<$. Employing non-reduced Heisenberg operators and the unitary evolution operator as in Eq. (1.7), they established

$$Z(\gamma_0)\text{X}(u) = Z(\gamma_f)\tilde{\text{X}}(-u+i\beta), \quad (2.35)$$

where the $Z$'s are the partition functions at the beginning and end of the process; thus $Z(\gamma_0) = \sum_{\gamma_0} e^{-\beta\varepsilon_{\gamma_0}} = e^{-\beta F_0^0}$ and similarly for $Z(\gamma_f)$. The relation (2.35) provides an alternate route to obtain the Crooks–Tasaki FT and the Jarzynski W-E relationship, see below.

We now show that the relation (2.35) can easily be obtained in a *theory with dissipation*, employing the Schrödinger forms. Starting with (2.28), we split $W_2$ in a conditional probability and the canonical distribution. Hence we have

$$Z(\gamma_0)\text{X}(u) = \iint \Delta\gamma_0\, \Delta\gamma_f\, P_{\text{ns}}(\gamma_f, t_f | \gamma_0, t_0)e^{-\beta\varepsilon_{\gamma_0}}e^{iu\varepsilon_{\gamma_f}}e^{-iu\varepsilon_{\gamma_0}}$$

$$= \iint \Delta\gamma_0\, \Delta\gamma_f\, P_{\text{ns}}(\gamma_f, t_f | \gamma_0, t_0)e^{-i(u-i\beta)\varepsilon_{\gamma_0}}e^{iu\varepsilon_{\gamma_f}}$$

$$= \iint \Delta\gamma_0\, \Delta\gamma_f\, \tilde{P}_{\text{ns}}(\gamma_0, t_0 | \gamma_f, t_f)e^{-\beta\varepsilon_{\gamma_f}}e^{i(u-i\beta)\varepsilon_{\gamma_f}}e^{-i(u-i\beta)\varepsilon_{\gamma_0}}$$

$$= \iint \Delta\gamma_0\, \Delta\gamma_f\, \tilde{P}_{\text{ns}}(\gamma_0, t_0 | \gamma_f, t_f)e^{-\beta\varepsilon_{\gamma_f}}e^{i(-u+i\beta)(\varepsilon_{\gamma_0}-\varepsilon_{\gamma_f})} = Z(\gamma_f)\tilde{\text{X}}(-u+i\beta). \quad (2.36)$$

The Fourier inversion of the left-hand side is $Z(\gamma_0)p(w)$, while for the right-hand side we need:

$$\frac{1}{2\pi}\int_{-\infty}^{\infty}du\,e^{-iuw}\int_{-\infty}^{\infty}dw'\,e^{-iw'(u-i\beta)}\tilde{p}(w')$$

$$=\frac{1}{2\pi}\int_{-\infty}^{\infty}dw'\,\tilde{p}(w')\int_{-\infty}^{\infty}du\,e^{-iu(w+w')}e^{-\beta w'}$$

$$=\int_{-\infty}^{\infty}dw'\tilde{p}(w')\delta(w+w')e^{-\beta w'}=e^{\beta w}\tilde{p}(-w). \qquad (2.37)$$

So we once more established the Crooks–Tasaki FT,

$$p(w)/\tilde{p}(-w)=[Z(\gamma_f)/Z(\gamma_0)]e^{\beta w}=e^{\beta(w-\Delta F^0)}, \qquad (2.38)$$

*Discussion.* In most of the standard literature all kinds of assumptions are made regarding the coupling to the reservoir. Most authors assume that at $t_0$ the system is coupled to the bath, after which it continues, being left isolated. Then at $t_f$ it is somehow reconnected 'without performing work', cf. [9, 42]. In a true quantum treatment, *whatever happens in the interval between the two time-points is irrelevant, or even inaccessible*! Note that the derivation in (2.36) is a purely mathematical exercise. This is in a sense similar to the quantum scattering problem. While in classical Rutherford scattering an $\alpha$-particle follows a hyperbolic path, in a quantum version the particle simply makes a transition $m\to n$, or — a particle with state $|m\rangle$ is annihilated and a particle with state $|n\rangle$ is created. Nothing else needs to be said. We should take quantum theory at face value!

## 2.2 Results based on the mesoscopic Master Equation for operators in a-space with coarse-grained states

When observables other than fields are varied generally, these observables do not commute with each other or with the Hamiltonian and a mesoscopic description is necessitated. Examples are shearing in an isothermal fluid or compression of a gas, as mentioned already in the introduction. Van Kampen [11] has outlined a procedure to coarse-grain such operators so that they quasi-commute by expressing them in the eigenstates $\{|\eta\rangle\}$ of the Hamiltonian setting,

$$a_k = \sum_{\eta,\eta'}|\eta\rangle\langle\eta|a_k|\eta'\rangle. \qquad (2.39)$$

Next, the states are grouped into energy cells $|\eta\rangle\in\Delta E_\kappa$ and the matrix elements between different energy cells are erased, so that

$$a_{k,cg}=\sum_\kappa\sum_{\eta,\eta'\in\Delta E_\kappa}|\eta\rangle\langle\eta|a_k|\eta'\rangle. \qquad (2.40)$$

For each $a_i$ a unitary transformation is made to diagonalize the matrix elements in each cell, hence

$$a_{i,cg}=\sum_\kappa\sum_{\bar{\eta}_i\in\Delta E_\kappa}|\bar{\eta}_i\rangle\langle\bar{\eta}_i|a_i|\bar{\eta}_i\rangle. \qquad (2.41)$$

Since the new sets of projectors all commute with each other and the Hamiltonian, the $\{a_{i,cg}\}$ have the mesoscopic eigenstates $\{|\alpha_\kappa\rangle\}$; the subscripts 'cg' on the $\{a_i\}$ will henceforth be omitted.

A mesoscopic ME can be derived, but is not needed here. However, we shall use the mesoscopic conditional probability

$$P(a_f, t_f | a_0, t_0) = P(\gamma_f | t_f | \gamma_0, t_0) \chi(a_f), \tag{2.42}$$

is the density of states; the symbol $a$ represents all the relevant variables that are being varied. We noted hereby that a given initial state $\gamma_0$ engenders a specific initial $a_0$, but the converse is not true. The mesoscopic conditional probability densities for forward and backward driving processes are therefore related by

$$P(a_f, t_f | a_0, t_0) = \tilde{P}(a_0, t_0 | a_f, t_f)[\chi(a_f)/\chi(a_0)]. \tag{2.41}$$

We now need the initial distributions for both sides. They are not simply the canonical distributions of the previous section, since the $a$'s are subject to additional fluctuations. In Ref. [12] we showed that here we need the canonical Boltzmann–Einstein probability density functions

$$W(a_0, t_0) = \hat{c}_0^{-1} e^{-\beta[F(a_0) - F_0^0]}, \quad W(a_f, t_1) = \hat{c}_f^{-1} e^{-\beta[F(a_f) - F_f^0]}, \tag{2.42}$$

where the non-superscripted $F$'s are nonequilibrium free energy functions, while the $\hat{c}$'s are normalization constants that vanish logarithmically.[6] Multiplying (2.41) with the $W$'s of (2.42), simple algebra yields

$$W_2(a_f, t_f; a_0, t_0)[\chi(a_0)\delta(a_0)/\chi(a_f)\delta(a_f)]e^{\beta[F(a_0)-F(a_f)]} = \tilde{W}_2(a_0, t_0; a_f, t_f)e^{-\beta \Delta F^0}. \tag{2.43}$$

Noting that $\chi(a)\delta a = \Delta \Gamma(a)$ is the accessible number of quantum states, whose logarithm is the nonequilibrium Gibbs entropy function, we finally obtain

$$W_2(a_f, t_f; a_0, t_0)e^{-\beta[E(a_f)-E(a_0)]} = \tilde{W}_2(a_0, t_0; a_1, t_1)e^{-\beta \Delta F^0}. \tag{2.44}$$

As for the microscopic case, we must compute the pdf for work from the characteristic function $\langle e^{iu\hat{W}} \rangle$, where basically $\hat{W}$ is still a course-grained operator with eigenvalues $\mathsf{W}$ (in some cases it could be treated as a classical variable, appealing to the correspondence limit). One obtains,

$$p(\mathsf{W}) = \iint da_f da_0 W_2(a_f, t_f; a_0, t_0) = \delta\left(\mathsf{W} - [E(a_f) - E(a_0)]\right), \tag{2.45}$$

as is also intuitively clear; a similar expression applies for $\tilde{p}(\mathsf{W})$. Multiplying (2.44) with the delta expressions and substituting the probability density functions for work, subsequent two-point integration yields the Crooks–Tasaki fluctuation theorem

---

[6] The $W(a)$ is given by the Boltzmann–Einstein principle: $\ln W(a) = -\beta[F(a) - F^0]$. Upon exponentiation we need a normalization constant, so that $W(a)\hat{c} = \exp\{-\beta[F(a) - F^0(a^0)]\}$. Normalizing, $\int W(a) da \approx W(a^0)\delta(a) = 1$, so that with $a \to a^0$ and $\exp \{ \} = 1$, we find $\ln \hat{c} \approx \ln \delta a = O(\ln N)$.

$$p(W)/\tilde{p}(-W) = \exp[\beta(W - \Delta F^0). \tag{2.46}$$

Multiplying by $\tilde{p}(-W)\exp(-\beta W)$ and integrating over W yields the Jarzynski relationship

$$\langle \exp(-\beta \hat{W}) \rangle = \exp(-\beta \Delta F^0). \tag{2.47}$$

Other properties for mesoscopic processes are found in Ref. [12].

## 3. Quantum Fluctuation Theorems for entropy-change and for entropy flow; asymptotic expressions

### 3.1 Isolated systems and microscopic considerations

It is well known that in any ensemble, microcanonical or general canonical, the Gibbs entropy is given by $S_G^0 = -k_B \operatorname{Tr} \rho \ln \rho$. For nonequilibrium ensembles, the Gibbs entropy function is generalized to be defined by

$$S_G(t) = -k_B \operatorname{Tr} \rho(t) \ln \rho(t). \tag{3.1}$$

We note that $S_G(t)$ is a smoothly varying scalar function of time, contrary to the Boltzmann entropy function. However, omitting the trace over $\rho$, we can define a stochastic entropy operator

$$\mathrm{S}(t) = -k_B \ln \rho(t), \tag{3.2}$$

which is a natural Schrödinger operator, in contrast with all other operators for FT's employed in this article. In the next section we shall likewise introduce an operator for entropy flow I to the environment or reservoir. In an isolated system the transition rates $W_{out} = W_{in}$, so that $\mathrm{I} = 0$, cf. footnote 7 in subsection 3.2. Under these circumstances we can employ a microscopic description, which unfortunately only applies to isolated systems. We are interested in the probability for the eigenvalues $\{\Delta s\}$ on the time interval of the forward protocol, which in this section we take to be $(t' \to t)$. This is most easily obtained from the generating function,

$$\Psi(\lambda) \equiv \langle e^{-\lambda[\mathrm{S}(t)-\mathrm{S}(t')]/k_B} \rangle_> = \langle e^{\lambda[\ln \rho(t) - \ln \rho(t')]/k_B} \rangle_>. \tag{3.3}$$

Now in the representation $\{|\gamma\rangle\}$ we have for the eigenvalues $\{\Delta s\}$

$$e^{-\lambda \Delta s/k_B} = \operatorname{Tr}\left\{e^{\lambda[\ln \rho(t) - \ln \rho(t')]/k_B}\right\} = \left(p(\gamma,t)/p(\gamma',t')\right)^\lambda. \tag{3.4}$$

Hence, doing the two-point averaging,

$$\Psi(\lambda) = \langle e^{-\lambda \Delta s/k_B} \rangle_> = \iint \Delta \gamma \, \Delta \gamma' \langle \left(p(\gamma,t)/p(\gamma',t')\right)^\lambda \rangle P_{\mathrm{ns}}(\gamma,t|\gamma',t') \, p(\gamma',t'). \tag{3.5}$$

Further, simple manipulation and time-reversal symmetry gives

$$\Psi(\lambda) = \iint \Delta\gamma\, \Delta\gamma'\, \left(p(\gamma',t')/p(\gamma,t)\right)^{-\lambda+1} \tilde{P}_{\text{ns}}(\gamma',t'|\gamma,t)\, p(\gamma,t)$$

$$= \langle e^{-(1-\lambda)\Delta s/k_B} \rangle_< = \tilde{\Psi}(1-\lambda)\,. \tag{3.6}$$

It remains to invert the generating functions. The left-hand side gives $p(\Delta s/k_B)$. For the right-hand side we have, using the inverse Laplace transform and setting $R \equiv \Delta s/k_B$

$$\int_{c-i\infty}^{c+i\infty} d\lambda\, e^{\lambda R} \int_0^\infty dR'\, \tilde{p}(R')\, e^{-(1-\lambda)R'}$$

$$= \int_0^\infty dR'\, e^{-R'}\, \tilde{p}(R') \int_{-i\infty}^{i\infty} d\lambda\, e^{\lambda(R+R')}$$

$$= \int_0^\infty dR'\, e^{-R'}\, \tilde{p}(R')\, \delta(R+R') = e^R \tilde{p}(-R)\,. \tag{3.7}$$

We thus established the quantum "transient entropy FT":

$$p(\Delta s) = e^{(\Delta s/k_B)} \tilde{p}(-\Delta s)\,. \tag{3.8}$$

The word transient stems from the classical description in which the stochasticity is attributed to the succession of jumps $w_{\sigma,\sigma'}(t)$ of the classical path.

Next, let us assume that the driven system has reached a steady state at some time $t^*$, from whereon the protocol will be time-independent, so that $p = \tilde{p}$. We then find the "steady-state entropy FT":

$$p(-\Delta s) = e^{-(\Delta s/k_B)} p(\Delta s)\,. \tag{3.9}$$

This relationship provides a quantitative answer to Loschmidt's objections to Boltzmann's irreversible H-theorem: decreasing entropy can be observed but with an exponentially low probability. An asymptotic form will be given later. Curiously, the results (3.8) and (3.9) will also be found for thermostatted systems, but a far more elaborate computation awaits!

### *3.2 Quantum entropy theorems for thermostatted systems based on the nonstationary mesoscopic master equation*

For thermostatted systems we need the nonstationary ME for mesoscopic states, since the system operators are randomized by the interactions $\lambda V$ with the reservoir. Reminiscent of the case of the spin lattice, with spin flips modeling the interactions, the mesoscopic states will not be denoted by $\{|\alpha\rangle\}$ as in Section 2.2 but by $\{|\sigma\rangle\}$, as is customary in the literature for entropy FT's. The ME reads [10], [43]:

$$\frac{\partial p(\sigma,t)}{\partial t} = \sum_{\sigma' \neq \sigma} \left[ w_{\sigma',\sigma}(t)p(\sigma',t) - w_{\sigma,\sigma'}(t)p(\sigma,t) \right] = -M_\sigma[p(\sigma,t)], \quad (3.10)$$

where $M_\sigma$ is the function-space master operator; further, $w_{\sigma',\sigma}(t)$ is the time-dependent transition rate *from $\sigma$ to $\sigma'$*; these rates are connected to the microscopic rates by

$$w_{\sigma,\sigma'}(t) = W_t(\gamma'|\gamma)\chi(\sigma'), \quad w_{\sigma',\sigma}(t) = W_t(\gamma|\gamma')\chi(\sigma), \quad (3.11)$$

where $\chi$ is the density of states. Because of microscopic reversibility for the $W$'s, we have

$$w_{\sigma,\sigma'}(t)/w_{\sigma',\sigma}(t) = \chi(\sigma')/\chi(\sigma). \quad (3.12)$$

We proceed to find the nonequilibrium Gibbs entropy function from the master equation. Computing $S_G$ in the representation $\{|\sigma\rangle\}$, we have

$$S_G(t) = -k_B \operatorname{Tr} \rho(t) \ln \rho(t) = -k_B \sum_\sigma p(\sigma,t) \ln p(\sigma,t). \quad (3.13)$$

From the ME (3.10) we find,

$$\frac{\partial S_G[p(\sigma,t)]}{\partial t} = \frac{1}{2}k_B \sum_{\sigma,\sigma'} [p(\sigma,t)w_{\sigma,\sigma'}(t) - p(\sigma',t)w_{\sigma',\sigma}(t)] \ln\left[\frac{p(\sigma,t)}{p(\sigma',t)}\right]$$

$$= k_B \sum_{\sigma,\sigma'} p(\sigma,t)w_{\sigma,\sigma'}(t) \ln\left[\frac{p(\sigma,t)}{p(\sigma',t)}\right]. \quad (3.14)$$

We now follow Schnakenberg [44] in splitting this into an entropy production rate $\eta$ and entropy current $I$ as follows:

$$\eta[p(\sigma,t)] = \frac{1}{2}k_B \sum_{\sigma,\sigma'} [p(\sigma,t)w_{\sigma,\sigma'}(t) - p(\sigma',t)w_{\sigma',\sigma}(t)] \ln\left(\frac{p(\sigma,t)w_{\sigma,\sigma'}(t)}{p(\sigma',t)w_{\sigma',\sigma}(t)}\right)$$

$$= k_B \sum_{\sigma,\sigma'} p(\sigma,t)w_{\sigma,\sigma'}(t) \ln\left(\frac{p(\sigma,t)w_{\sigma,\sigma'}(t)}{p(\sigma',t)w_{\sigma',\sigma}(t)}\right), \quad (3.15)$$

$$I_\eta[p(\sigma,t)] = k_B \sum_\sigma p(\sigma,t) \sum_{\sigma'} w_{\sigma,\sigma'}(t) \ln\left(\frac{w_{\sigma,\sigma'}(t)}{w_{\sigma',\sigma}(t)}\right). \quad (3.16)$$

This gives the entropy conservation rule,

$$\partial S_G/\partial t + I_\eta = \eta \geq 0, \quad (3.17)$$

by Klein's lemma, cf. the right-hand side of the first line of (3.15).[7]

All these expressions are still averages, as implied by the sum $\sum_\sigma p(\sigma,t)[\ldots]$. Omitting this sum, we have the corresponding fluctuating quantities being the $\sigma$-representation of operators $\hat{S}$, $\hat{I}$ and $\hat{\eta}$. We will also set $\hat{I} = \partial \Delta S_{flow}/\partial t$, the total entropy change being

---

[7] When $I = 0$, the $\chi$'s of (3.11) are unity, so the system can be handled with the microscopic states $\{|\gamma\rangle\}$.

$$\mathrm{Tr}\rho[\Delta \mathrm{S}_{total}] = \Delta s_{syst} + \Delta s_{flow}. \quad (3.18)$$

*Computation for $\Delta s_{flow}$.*

Various authors have invented elaborate schemes to obtain $\Delta s_{flow}$ — like the "quantum Hamiltonian" method by Harris and Schütz [23] in which projectors $|\sigma\rangle\langle\sigma|$ and 'pseudo-projectors' $|\sigma\rangle\langle\sigma'|$ are employed. Here we shall give a direct computation with no artificial concepts. Like for the work FT's, it does not suffice to work with the ME for the scalar probabilities. Rather, we need the operator form in the Liouville space associated with the tensor product space $\bar{\mathbf{S}}[\mathrm{H}^{cg}(t)] \otimes \bar{\mathbf{S}}[\mathrm{H}^{cg}(t')]$, where $\bar{\mathbf{S}}[\mathrm{H}^{cg}(t)]$ and $\bar{\mathbf{S}}[\mathrm{H}^{cg}(t')]$ are the Hilbert spaces for the coarse-grained Hamiltonians $\mathrm{H}^{cg}(t)$ and $\mathrm{H}^{cg}(t')$, respectively. We can then convert to a Schrödinger form, similarly as in our previous developments.

Let $\mathrm{S}_{flow}(0)$ be the Schrödinger operator for the entropy flow at $t' = 0$, the beginning of the forward protocol and let $\mathrm{S}^{R}_{H,flow}(t)$ be the reduced Heisenberg operator for the entropy flow at time $t$. Then for the generating function of the eigenvalues $\Delta s_{flow}$ we have in the indicated Liouville space

$$\Psi_{\Delta s_{flow}}(\lambda, t) = \left\langle \exp\{-\lambda[\mathrm{S}^{R}_{H,flow}(t) - \mathrm{S}_{flow}(0)]\} \right\rangle$$

$$= \left\langle \exp[-\lambda \mathrm{S}^{R}_{H,flow}(t)] \exp[\lambda \mathrm{S}_{flow}(0)] \right\rangle. \quad (3.19)$$

Writing $\mathrm{S}^{R}_{H,flow}$ in its projectors, we have

$$-\lambda \mathrm{S}^{R}_{H,flow}(t) = \left\{ -\lambda \mathrm{T} \exp\left( -\sum_{\sigma\sigma'} |\sigma\sigma'\rangle\langle\sigma'\sigma| \int_0^t d\vartheta M_\sigma(\vartheta) \right) \right\} \mathrm{S}_{flow}(t)$$

$$= -\lambda \sum_{\sigma\sigma'} |\sigma\sigma'\rangle\langle\sigma'\sigma| \left\{ \mathrm{T} \exp[-\int_0^t d\vartheta M_\sigma(\vartheta)] \right\} \mathrm{S}_{flow}(t), \quad (3.20)$$

where we used an extension of footnote 4. Now for the matrix elements of the entropy flow operators we have as usual,

$$\langle\sigma'\sigma | \mathrm{S}_{flow}(t) | \sigma\sigma'\rangle = \langle\sigma | \mathrm{S}_{flow}(t) | \sigma\rangle, \quad (3.21)$$

$$\langle\sigma' \quad \mathrm{S}_f \qquad \qquad \mathrm{S} \; \sigma|_{\;l}$$

$$(3.22)$$

Taking the scalar product in (3.20) and substituting these results into the above, we obtain

$$\Psi_{\Delta s_{flow}}(\lambda, t) = \left\langle \exp\left\{ -\lambda \mathrm{T} \left[ \exp\left( -\int_0^t d\vartheta M_\sigma(\vartheta) \right) \right] \langle\sigma | \mathrm{S}_{flow}(t) | \sigma\rangle \langle\sigma' | \exp\lambda \mathrm{S}_{flow}(0) | \sigma'\rangle \right\} \right\rangle. \quad (3.23)$$

We should remember that the states $\{|\sigma'\rangle\}$ at $t = 0$ and the states $\{|\sigma\rangle\}$ at $t$ belong to different Hamiltonians; more about that later in the discussion. Although $\mathrm{S}_{flow}(t)$ is not an observable in this state space, the generating function should exist for most $\lambda$. Thus, $\Psi_{\Delta s_{flow}}(\lambda, t)$ is a correlation function $\Phi_{\mathrm{AB}}(t,0)$ such as studied in section 2, with

$$\langle\sigma|A_{Hd}^{R}(t)|\sigma\rangle = \exp\left\{-\lambda T \exp\left[\int_0^t d\vartheta M_\sigma(\vartheta)\right]\langle\sigma|S_{flow}(t)|\sigma\rangle\right\}, \tag{3.24}$$

while

$$\langle\sigma'|B_d(0)|\sigma'\rangle = \exp(-\lambda\langle\sigma'|S_{flow}(0)|\sigma'\rangle). \tag{3.25}$$

For expediency we set $\langle\sigma'|S_{flow}(0)|\sigma'\rangle = 0$. Multiplying and averaging – i.e. summing over the two states relating to the two-point measurement, $\sum_{\sigma,\sigma'}$ – from (2.27) we find the Schrödinger form; denoting the eigenvalue of $S_{flow}(t)$ by $\Delta s_{flow}$ and with $-\lambda\Delta s_{flow} = -\lambda\int_0^t d\vartheta I_\sigma(\vartheta)$, we obtain

$$\Psi_{\Delta s_{flow}}(\lambda,t) = \sum_{\sigma\sigma'} P_{ns}(\sigma,t|\sigma',0)p(\sigma') T \exp\left(-\lambda\int_0^t d\vartheta w_{\sigma\sigma'}(\vartheta)\ln\left[\frac{w_{\sigma\sigma'}(\vartheta)}{w_{\sigma'\sigma}(\vartheta)}\right]\right). \tag{3.26}$$

*This is the main result.* It looks intuitively 'plausible' and is by and large quite similar to the corresponding expression in the long paper by Harris and Schütz, yet is far less complex, cf. [23, Eq. (3.21)].[8] However, no assumptions about jumps, giving rise to stochastic trajectories have been made anywhere. Rather, as for the work FT's, we computed entropy flow as a reduced Heisenberg correlation function and then, via differentiation implied, obtained a tractable Schrödinger form.

Now the following manipulations are done to connect with the reverse generating function:

$$T \exp\left\{-\lambda\int_0^t d\vartheta w_{\sigma,\sigma'}(\vartheta)\ln\left(w_{\sigma,\sigma'}(\vartheta)/w_{\sigma',\sigma}(\vartheta)\right)\right\}$$

$$= T \exp\left\{(1-\lambda)\int_0^t d\vartheta w_{\sigma',\sigma}(\vartheta)\ln\left(w_{\sigma,\sigma'}(\vartheta)/w_{\sigma',\sigma}(\vartheta)\right)\right\}$$

$$= T \exp\left\{-(1-\lambda)\int_0^t d\vartheta w_{\sigma',\sigma}(\vartheta)\ln\left(w_{\sigma',\sigma}(\vartheta)/w_{\sigma,\sigma'}(\vartheta)\right)\right\}$$

$$= T \exp\left\{-(1-\lambda)\int_0^t d\vartheta w_{\sigma',\sigma}(t-\vartheta)\ln\left(w_{\sigma',\sigma}(t-\vartheta)/w_{\sigma,\sigma'}(t-\vartheta)\right)\right\}, \tag{3.27}$$

where in the last transition we used that $\int_0^t d\vartheta f(\vartheta) = \int_0^t d\vartheta f(t-\vartheta)$. Put into (3.26), we showed

$$\left\langle e^{-\lambda\Delta s_{flow}}\right\rangle_> = \left\langle e^{-(1-\lambda)\Delta s_{flow}}\right\rangle_<. \tag{3.28}$$

This is the same result as for the isolated system, cf. (3.6). Needless to say, that Eq. (3.6) is also applicable to the system entropy of the thermostatted system, providing we set $\lambda \Rightarrow \sigma$. Since the two entropies of (3.18) are additive, their generating functions are multiplicative, except that we must avoid 'double averaging'. For the purist we can also write down the generating function for the total entropy change immediately, by putting (3.4) inside (3.26) and summing over $p(\sigma',0)P$:

---

[8] With $\tilde{H}_Q$ being their 'quantum Hamiltonian', the generating function $\langle e^{-\lambda R}\rangle$ has the form $\exp[-\int d\vartheta \tilde{H}_Q(\lambda,\vartheta)$ $= -\exp\left(-\int d\vartheta w_{\sigma,\sigma'}(\vartheta)\{\exp[-\lambda\ln(w_{\sigma,\sigma'}/w_{\sigma',\sigma})]\}\right)$. The $\lambda$ should clearly be in the lower exponential to match the generating function. With that change the result simplifies considerably and matches in form our result, Eq. (3.26).

$$\left\langle e^{-\lambda \Delta s_{total}} \right\rangle_>$$

$$= \sum_{\sigma\sigma'} [p(\sigma,t)]^\lambda \, \text{T} \exp\left\{-\lambda \int_0^t d\vartheta \, w_{\sigma,\sigma'}(\vartheta) \ln\left(\frac{w_{\sigma,\sigma'}(\vartheta)}{w_{\sigma',\sigma}(\vartheta)}\right)\right\} [p(\sigma',0)]^{1-\lambda} P_{ns}(\sigma,t \mid \sigma',0). \quad (3.29)$$

The steps of (3.27) can be verbatim repeated, using in addition time-reversal symmetry for $P_{ns}$. Hence, again we arrive at

$$\left\langle e^{-\lambda \Delta s_{total}} \right\rangle_> = \left\langle e^{-(1-\lambda)\Delta s_{total}} \right\rangle_<. \quad (3.30)$$

The inversion of this result goes as in (3.7), giving the "transient entropy FT",

$$p(\Delta s_{total}) = e^{(\Delta s_{total}/k_B)} \tilde{p}(-\Delta s_{total}). \quad (3.31)$$

If a steady state is reached at some point, then with $p = \tilde{p}$, we find the steady state entropy FT, giving the probability that entropy decrease occurs,

$$p(-\Delta s_{total}) = e^{-(\Delta s_{total}/k_B)} p(\Delta s_{total}). \quad (3.32)$$

Asymptotically, $\Delta s_{total} \sim \eta t$, where $\eta$ is the entropy production rate as defined in (3.15). The asymptotic FT now reads $p(-\eta) \sim e^{-\eta t} p(\eta)$; or also

$$\lim_{t \to \infty} \left( -\frac{1}{t} \ln\left[\frac{p(-\eta)}{p(\eta)}\right] \right) = \eta. \quad (3.33)$$

*Further discussion*

The entropy flow current is based on the quantity $w_{\sigma\sigma'}(t)$, which we used cavalierly but we indicated along the way that $\sigma$ and $\sigma'$ are ES of different Hamiltonians and, surely, this rate could not represent a single transition! Even less could it be obtained from Fermi's 'golden rule'. Fortunately, the final result led to a sum of time-ordered convolution integrals, Eq. (3.26). This alone indicates that multiple transitions are involved. More realistically, the result should have been obtained with the segmentation method. If the points $t_1, t_2, \ldots t_{n-1}$ are spaced close enough, integration means multiplication with the interval $t_{i+1} - t_i$; in that case we would have obtained

$$\text{T} \exp\left(-\lambda \int_0^t d\vartheta \, w_{\sigma\sigma'}(\vartheta) \ln\left[\frac{w_{\sigma\sigma'}(\vartheta)}{w_{\sigma'\sigma}(\vartheta)}\right]\right) = \lim_{n \to \infty} \exp\left(-\lambda \sum_{i=0}^{n-1} w_{\sigma_i \sigma_{i+1}}(t_{i+1} - t_i) \ln\left[\frac{w_{\sigma_i \sigma_{i+1}}(t_i)}{w_{\sigma_{i+1} \sigma_i}(t_i)}\right]\right). \quad (3.34)$$

Or, we can write this as a product of "nearest neighbour" correlations,

$$\text{T} \exp\left(-\lambda \int_0^t d\vartheta\, w_{\sigma\sigma'}(\vartheta) \ln\left[\frac{w_{\sigma\sigma'}(\vartheta)}{w_{\sigma'\sigma}(\vartheta)}\right]\right) = \lim_{n\to\infty} \prod_{i=0}^{n-1} \exp\left(-\lambda\, w_{\sigma_i\sigma_{i+1}}(t_{i+1}-t_i)\ln\left[\frac{w_{\sigma_i\sigma_{i+1}}(t_i)}{w_{\sigma_{i+1}\sigma_i}(t_i)}\right]\right).$$
(3.35)

[Actually, the number of transitions should remain finite and of the order of *N*, where *N* is obtained as the time interval of the protocol divided by the duration of a microscopic transition.]

This is the result in terms of single transitions; the *symbolic* overall result is the convolution integral at the left-hand sides.

## 4. Conclusions

In section 1 we presented a brief survey of previous work and some of our main objections. The oldest articles on FT's, particularly the entropy fluctuation theorem, were all of a classical nature. While the volume of an assembly of phase points is a Poincaré invariant in an equilibrium state, in nonequilibrium the volume generally contracts. The ratio of probabilities for entropy producing to entropy reducing processes can then be computed from the spectrum of Lyapunov exponents, as by Evans and Searles, Evans, Cohen and Morriss, and others quoted previously. While the theory of chaos, formulated mainly after the nineteen-sixties, is extremely interesting as a branch of mathematics, it still is a *classical* phase-space description, which in this author's view violates the very tenets of quantum mechanics and quantum statistics. Surely, such approaches can give the 'right' results, but for the wrong reasons; another example is afforded by Rutherford scattering of alpha-particles in the early twentieth century, in which the assumed hyperbolae of classical mechanics are purely illusory.

More interesting are quantum approaches that use the von Neumann equation or Heisenberg operators to describe such observables as *work*, or *entropy-flow* that require to be defined by generating functions or characteristic functions, since the observable involved is not part of the Hilbert space of the leading Hamiltonian $H^0$. These theories may produce beautiful results, such as the Talkner–Hänggi relationship for the characteristic function of work. Unfortunately, 'pure' Heisenberg operators such as used in Kubo's original LRT have non-convergent correlation functions — e.g., the mixing theorem does not hold for $t \to \infty$ — and show no dissipative behaviour or entropy production, despite yielding a formal *fluctuation–dissipation theorem*! Amended Kubo theory, with convergent results has been the research area of the author and coworkers since the mid-seventies, leading to '*Reduced Heisenberg operators*' in which the coupling $\lambda V$ with an external reservoir or internal causes has been carried out to all orders of perturbation. The initial research is described in articles [1-3] published in J. Math. Phys. (1978-1982).

Now let us move on to the present quest involving FT's for driven systems with time-dependent Hamiltonians, some forty years later. The time dependence is generally thwarted by considering piece-wise Hamiltonians over short intervals of the protocol; the 'segmentation method' is described at the end of our introductory Section 1. Yet in this article, contrary to our previous paper [10], it is not explicitly used, an exception being in the discussion notes of 3.2.

Surprisingly, the only Schrödinger operator that presents itself naturally is the Gibbs' entropy. It is well known that in any ensemble we have $S_G = -k_B \text{Tr}\,\rho\ln\rho$, where $\rho$ is the density operator. The stochastic equivalent is the operator $S(t) = -k_B \ln\rho(t)$. Unfortunately, this form can only be used for an isolated system in which there are no other entropies like entropy flow to the reservoir.

In all other cases we deal with Heisenberg operators, as in Kubo's LRT. They are usually outside the state space under consideration, but, as pointed out by von Neumann, their generating function or the characteristic function, $\Psi(\lambda)$ or $X(u)$, respectively, exist within a wide domain of $\lambda$ or $u$. For FT's we need the reduced operators' diagonal parts, which are responsible for the dissipation. Generally this part goes as $\text{T}\exp[-\int_0^t d\vartheta\,\Lambda(\vartheta)]$ where $\Lambda$ is positive definite (allowing for convergence of correlation and response functions) and has a spectral resolution in terms of projectors and the master equation in function-space. Reduced Heisenberg operators are therefore by nature stochastic. Now comes the "crux": *we must convert these operators and their correlation functions to the Schrödinger form*! This is accomplished by differentiation with initial conditions, and then solving with a Green's function procedure. This was first shown by the author in Ref. [1, subsection 9.2 and previous]. In the Schrödinger form, the operators are time-independent, the time dependence now being vested in the conditional probability $P_{\text{ns}}(\gamma,t|\gamma',0)$ or in the two-point probability $W_2(\gamma_f, t_f; \gamma_0, t_0)$. The main equations for this procedure are found in (2.27) and (2.28). (There is also a version of this procedure in our book, [4, subsection 16.14.1].)

With the Schrödinger form we can now re-establish all results done previously with non-convergent Kubo theory. In particular, the Crooks–Tasaki work FT's are rapidly established, as well as the Jarzynski W–E relationship. Likewise, the Talkner–Hänggi relation for the symmetry properties of the characteristic function tumbles out effortlessly. Finally, we used this approach to ascertain the entropy-flow generating function and the subsequent symmetry properties that result in the 'total' entropy fluctuation theorem, all being in accord with results found by others (except a small variation from the Harris–Schütz result as published). All this is accomplished without assuming that there are 'stochastic trajectories'; in fact, what happens between initial and final measurement is ontologically inaccessible.

## Acknowledgements

This article was prepared as an invited presentation for the 11-13 July 2016 Solvay Workshop on "Nonequilibrium and nonlinear phenomena in statistical mechanics" at the Université Libre de Bruxelles. In particular, we are indebted to Madame Isabelle Van Geet for all her detailed arrangements. This article was submitted posthumously with edits by the author's grandson, Dr. David S. Sukhdeo.